\documentclass[
  aps,
  pra,
  superscriptaddress,
  amsmath,amssymb,
  floatfix,
  reprint,   
]{revtex4-2}

\usepackage{graphicx}
\usepackage{dcolumn}
\usepackage{bm}
\usepackage{braket}
\usepackage{xcolor}
\usepackage{soul}
\usepackage{csquotes}
\def\ANU{Center of Excellence for Quantum Computation and Communication Technology, The Department of Quantum Science and Technology, Research School of Physics, The Australian National University, Canberra, Australia}
\def\NUS{Centre for Quantum Technologies, National University of Singapore,
3 Science Drive 2, Singapore, Republic of Singapore}
\def\QINC{Quantum Innovation Centre (Q.InC), Agency for Science Technology and Research (A*STAR), 2 Fusionopolis Way, Innovis \#08-03, Singapore 138634, Republic of Singapore}

\begin{document}
\preprint{APS/123-QED}

\newcommand{\ZJ}[1]{\textcolor{blue}{#1}}
\newcommand{\AW}[1]{\textcolor{green}{#1}}

\title{All-Gaussian State Discrimination Beyond the Coherent Helstrom Bound}

\author{Angus Walsh}
\email{Angus.Walsh@anu.edu.au}
\affiliation{\ANU}
\author{Lorc\'an O. Conlon}
\email{lorcanconlon@gmail.com}
\affiliation{\QINC}
\affiliation{\NUS}
\author{Biveen Shajilal}
\affiliation{\QINC}
\author{\"Ozlem Erk\i l\i \c{c}}
\affiliation{\ANU}
\author{Jiri Janousek}
\affiliation{\ANU}
\author{Syed M. Assad}
\affiliation{\QINC}
\author{Jie Zhao}
\affiliation{\ANU}
\author{Ping Koy Lam}
\affiliation{\QINC}
\affiliation{\NUS}

\begin{abstract}
A core problem in communications is the optimal discrimination of binary-phase-shift-keyed (BPSK) signals. A longstanding goal has been to reach the fundamental quantum limit, known as the Helstrom bound, for BPSK signals encoded in coherent states. However, due to technical constraints, proposals for reaching the bound remain impractical. In this letter we take an alternative approach: using only Gaussian optics - displaced squeezed states and homodyne detection - we achieve discrimination of BPSK signals with error rates below what can be achieved using coherent states and any quantum measurement.
\end{abstract}

\maketitle


The theory of quantum state discrimination provides tools to study the fundamental limits of communication systems \cite{holevo1973statistical,barnett2009quantum}. Notably, the Helstrom bound specifies the minimum probability of error in distinguishing non-orthogonal signals \cite{helstrom1969quantum}. For a BPSK signal encoded in coherent states the Helstrom bound shows that a suitably designed receiver can surpass the standard quantum limit (SQL) for discriminating this signal. 
\par 
A number of receiver designs have been proposed that approach, or in some limits attain, the Helstrom bound for the coherent BPSK (C-BPSK) signal \cite{burenkov2021practical}. Kennedy introduced the idea of a receiver that displaces the incoming signal prior to photon counting, and in doing so surpasses the SQL \cite{kennedy1973near}. Variations on this theme have been able to improve upon its performance, and in some cases theoretically achieve the coherent Helstrom bound \cite{dolinar1973optimum,PhysRevA.71.022318,PhysRevLett.101.210501,PhysRevLett.117.200501}.
\par
Displacement-based receivers have several problems with their implementations. It is implicitly assumed that the strength of the received signal is known perfectly, however if the amplitude varies with time - for example due to fluctuations in the channel loss - then the displacements in the receiver will be unoptimised leading to excess errors. Receiver designs that attain the coherent Helstrom bound require precise real-time adjustment of the displacement, and may only reach the bound in an asymptotic limit. Additionally, the requirement for single-photon or photon-number-resolving detectors makes these designs sensitive to detector inefficiencies and dark noise, leading to errors that can proliferate in multi-stage receivers \cite{PhysRevLett.106.250503,PhysRevLett.121.023603,PhysRevA.101.032306}.
\par
The toolbox of Gaussian optics provides an experimentally accessible alternative to these approaches \cite{BachorRalph,RevModPhys.84.621,RevModPhys.77.513}. In contrast to single-photon detectors, homodyne detectors can be constructed with near-unity efficiency and large electronic noise clearance \cite{PhysRevLett.117.110801}. Recent advances in quadrature squeezing have demonstrated high levels of noise reduction and long-term stability \cite{PhysRevLett.117.110801,shajilal202212}, motivating the exploration of state discrimination using only Gaussian states and operations.
\par
Rather than constructing an elaborate receiver, the states in which the BPSK signal is encoded can be optimised instead. By injecting quadrature squeezed light into the communications channel we can obtain a quadratic improvement in the signal-to-noise ratio (SNR) of the BPSK signal for the same mean energy of the states. This makes it possible to surpass the Helstrom bound for the C-BPSK signal using only Homodyne detection \cite{PhysRevA.64.014304,PhysRevA.97.032315}. In this manuscript we demonstrate this experimentally for the first time, and outline the importance of this task for quantum communication.

\noindent {\it Theoretical Framework.—}In optical communications the C-BPSK signal consists of symbols with equal amplitude and opposite phase, represented by the quantum states 
\begin{equation}
\ket{\psi_k}=\hat{R}(k\pi)\hat{D}(\alpha)\ket{0}, k\in\{0,1\}
\end{equation}
where $\hat{D}$ and $\hat{R}$ are the displacement and rotation operators respectively. The coherent amplitude $\alpha$ is assumed to be real-valued. In a standard protocol these symbols are generated by a transmitter station and sent through an appropriate channel to a receiver station. The channel modifies the signal through the introduction of noise and optical losses. The performance of the  protocol is quantified by the probability of error in estimating the transmitted symbol,
\begin{align}
    P_{e} = P(0)P(1|0) +  P(1)P(0|1),
\end{align}
where $P(k)$ is the probability that the $k$ state was sent, and $P(j|k)$ the probability that our decision strategy mistakes the $k$ state for the $j$ state. We aim to minimise this error for a given mean photon number, $\bar{n} = \braket{\hat{N}}$, of the signal states. We can simplify the error probability by assuming that each symbol is \emph{a priori} equally likely - as this maximises the mutual information between the transmitter and receiver - then, noting that the BPSK signal is symmetric, we can write the probability of error as 
\begin{align}
    P_{e} = P(1|0) = P(0|1)
\end{align}
\par The probability of error for a specific protocol is a function of the signal-to-noise ratio (SNR) of the received signal. For Gaussian states, it is convenient to use the homodyne SNR, defined as
\begin{align}
    \text{SNR} := \frac{\braket{\hat{X}}^2}{\braket{\Delta\hat{X}^2}},
\end{align}
where $\hat{X}$ is a quadrature operator \cite{loudon2000quantum}. Coherent states have a SNR of $4\bar{n}$, however it is possible through quadrature squeezing to achieve a higher SNR for the same $\bar{n}$. Encoding a BPSK signal onto squeezed vacuum results in the displaced-squeezed-state BSPK (S-BPSK) signal
\begin{align}
    \ket{\psi_k} =\hat{R}(k\pi)\hat{D}(\beta)\hat{S}(r)\ket{0}, k\in\{0,1\},
    \label{DSS}
\end{align}
where $\hat{S}$ is the squeezing operator. The parameter $r$ is real-valued and positive so as to obtain amplitude squeezing, and we set the coherent amplitude $\beta = \sqrt{\alpha^2 - \sinh^2(r)}$ so that the S-BPSK and C-BPSK signals have the same energy and can justifiably be compared \footnote{For $\beta$ to be real-valued we require $\alpha^2>\sinh{^2(r)}$, in other words $\gamma<\frac{1}{2}$.}. For states of a given mean photon number there is an optimal distribution of energy between squeezing and displacement that maximises the SNR. We can define the squeezing ratio as the contribution of squeezing to the total energy of the state, $\gamma = \sinh^2(r)/\bar{n}$. When the ratio is equal to
\begin{align}
    \gamma = \frac{\bar{n}}{2\bar{n}+1}
\end{align}
a maximum SNR of $4(\bar{n}^2+\bar{n})$ is obtained, this being the maximum SNR for a Gaussian state. The introduction of squeezing therefore leads to quadratic scaling of the SNR over the coherent signal \cite{PhysRevA.64.014304,PhysRevA.97.032315}.
\par Using an optimal decision strategy the minimal probability of error when using a homodyne detector as a receiver is 
\begin{align}
    P_{\text{HOM}} = \frac{1}{2}\text{Erfc}[\sqrt{\frac{\text{SNR}}{2}}].
\end{align}
The fundamental lower limit on the discrimination of any two pure states, $\ket{\psi_0}$ and $\ket{\psi_1}$, is given by the Helstrom bound $P_{\text{HEL}} = \frac{1}{2}(1-\sqrt{1-|\braket{\psi_0|\psi_1}|^2})$ \cite{helstrom1969quantum}.
Using the fact that for any Gaussian BPSK signal $|\braket{\psi_0|\psi_1}|^2 = e^{-\text{SNR}}$ we can write the Helstrom bound as a function of the SNR
\begin{align}
    P_{\text{HEL}} = \frac{1}{2}(1-\sqrt{1-e^{-\text{SNR}}}),
\end{align}
which allows for direct comparison to the homodyne limit.
Table 1
\begin{table}[]
\begin{ruledtabular}
\begin{tabular}{c | c | c}
   Encoding & Homodyne Error & Helstrom Error \\
   \hline
   Coherent & $\frac{1}{2}\text{Erfc}[\sqrt{2\bar{n}}]$ & $\frac{1}{2}(1-\sqrt{1-e^{-4\bar{n}}})$ \\
   Squeezed & $\frac{1}{2}\text{Erfc}[\sqrt{2(\bar{n}^2+\bar{n})}]$ & $\frac{1}{2}(1-\sqrt{1-e^{-4(\bar{n}^2+\bar{n})}})$
\end{tabular}
\end{ruledtabular}
    \caption{Error bounds for BPSK communication protocols as functions of the mean photon number of the signal states. The signal is encoded either in coherent states or displaced squeezed states, and received using either a Homodyne detector or a detector that saturates the Helstrom bound.}
    \label{tab:my_label}
\end{table}
shows the homodyne limit and Helstrom bound for both the C-BPSK and S-BPSK signals. The coherent-homodyne limit is commonly referred to as the standard quantum limit (SQL). The squeezed-homodyne limit is the lowest achievable error rate using only Gaussian states and operations. Despite this constraint, due to the quadratic scaling of the displaced squeezed state SNR, this bound can fall below the coherent Helstrom bound for signals with the same mean photon number. \par
By applying another squeezing operator, $\hat{S}(-r)$, to the S-BPSK signal at the receiver we can obtain a coherent state with the same SNR. Therefore any receiver that saturates the Helstrom bound for coherent states can be made to saturate the bound for squeezed states simply by including this operator as a preamplifier. However, it is important to note that the definition of the S-BPSK signal in Equation (\ref{DSS}) is highly idealised. In practice the signal states will not be pure, due to excess noise in the anti-squeezed quadrature, and losses in the implementation of the displacement operation. Instead of squeezed vacuum the BPSK signal will be encoded onto squeezed thermal states characterised by a mean thermal photon number $\bar{n}_{th}$ which effectively represents wasted energy. Consequently the pure state Helstrom bound shown in Table (1) will not be applicable to any realistic scenario.


\begin{figure}
    \centering
    \includegraphics[width = \linewidth]{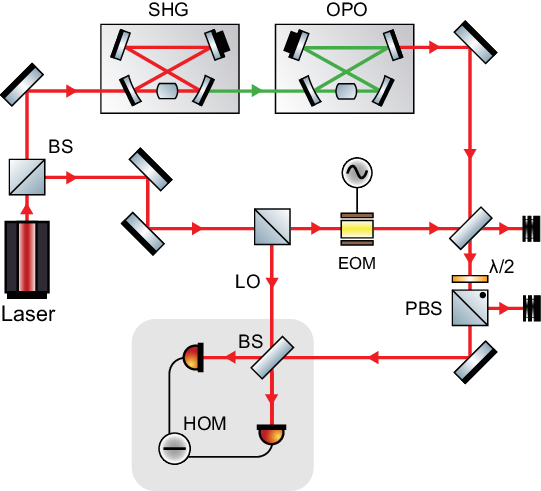}
    \label{fig:1}
    \caption{Experimental setup for all-Gaussian state discrimination of BPSK signals. A continuous-wave 1550 nm laser output is separated into a pump, signal and local oscillator (LO) by a network of beamsplitters (BS). The pump beam drives a second harmonic generation (SHG) cavity to produce light at $775$ nm which in turn pumps an optical parametric oscillator (OPO) to generate squeezed vacuum at $1550$ nm. Both cavities use periodically-poled potassium titanyl phosphate as a non-linear medium. An electro-optic modulator (EOM) displaces the $3$ MHz sideband of the signal beam which is then mixed with the OPO output on a biased BS to make the S-BPSK signal. Channel loss is implemented using a half-wave plate ($\lambda/2$) and a polarising beamsplitter (PBS) before the signal is measured at a homodyne receiver (HOM).}
\end{figure}
\noindent {\it Experimental demonstration.—}\mbox{Figure 1} shows the setup of our experiment. Using a 1550 nm laser source we use an SHG cavity to produce light at 775 nm, which in turn pumps an optical parametric oscillator (OPO) to generate squeezed vacuum at 1550 nm. An electro-optic phase modulator is used to generate coherent states at the 3 MHz sideband of a 1550 nm carrier beam, which are then mixed with the squeezed vacuum on a highly biased beam splitter (nominal $R = 99.5$) to produce displaced squeezed states. A lossy channel is simulated by using a half wave plate and polarising beam splitter, and after passing through the channel the signal is measured on a homodyne detector with visibility $V = (98.75\pm0.4)\%$. More detail on how other noise sources, including impure squeezing are included in Appendix~\ref{sec:imperfections}. A pair of acousto-optic modulators were used to produce a single sideband modulation at 20 MHz (not shown in the figure) which is injected into the OPO and co-propagates with the squeezed light, this serves as a common reference to lock the phases of the OPO pump, the coherent displacement, and the homodyne local oscillator (LO) \cite{shajilal202212}. In each run of the experiment the squeezing level is fixed and the coherent displacement is varied by increasing the amplitude of the 3 MHz modulation.

After demodulation of the 3 MHz sideband the homodyne current is recorded using an oscilloscope. With all the light blocked we first measure the electronic noise $v_{en}$ of the Homodyne detector, we then measure the shot noise from the LO to which all other measurements are normalised. The electronic noise is approximately 16 dB below shot noise. With the displacement beam blocked we measure the squeezing and anti-squeezing quadratures of the OPO output and calculate the squeezing parameter $r$ and thermal noise $v_{th}$. Finally the displacement beam is unblocked and the modulation signal in incrementally increased, varying $\beta$ while keeping the other parameters fixed. The measured state has an estimated mean photon number 
\begin{equation}
    \bar{n} = \frac{\beta^2 + \sinh^2(r) + (v_{th}-1)/2}{V^2}, \label{etot}
\end{equation}
from which the probability of error can be predicted by solving for $\beta$ and substituting into the Homodyne error
\begin{align}
    P_e = \frac{1}{2}\text{Erfc}[\sqrt{\frac{2(\bar{n}V^2-\sinh^2(r)-(v_{th}-1)/2)}{(e^{-2r}v_{th}+v_{en})}}], \label{expmod}
\end{align}
as shown in the dashed traces in \mbox{Figure 2 (a)}. To directly estimate the probability of error from the data we simply calculate the fraction of samples that are above (below) $0$ when the $\ket{\psi_1}$ ($\ket{\psi_0}$) state is transmitted. To quantify the uncertainty in out estimate of the error rate we treat each sample as a Bernoulli trial, and using the Jeffrey's prior calculate the posterior distribution for $P_e$. Due to the large number of samples for each state ($ n_s =5\times10^6$) the resulting Beta distribution is to a good approximation normal, and its standard deviation is shown as error bars.

\begin{figure*}
    \centering
    \includegraphics[width=1\linewidth]{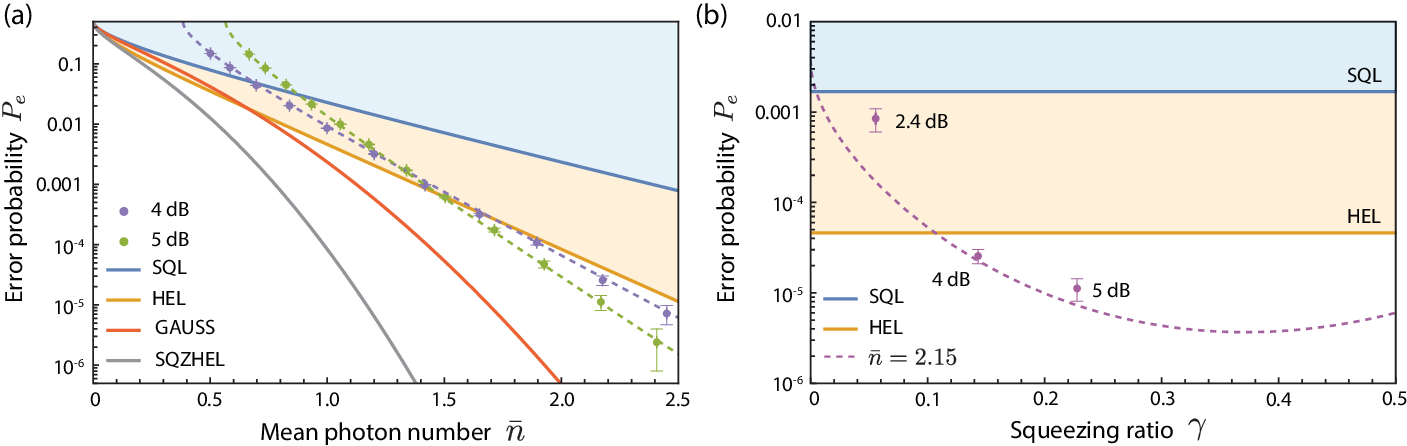}
    \caption{(a) Error probability in BPSK communication against the mean photon number of the signal. (b) Error probability in BPSK communication against the proportion of energy allocated to squeezing the signal, for states with mean photon numbers close to $\bar{n}=2.15$. Markers show experimental results from our implementation of a BPSK protocol with displaced squeezed states and homodyne detection. Dashed curves show the expected error given the experimental parameters, and solid curves represent the theoretical limits: the standard quantum limit of coherent states and homodyne detection (SQL), the Helstrom bound for coherent states (HEL), the limit of Gaussian states and measurements (GAUSS), and the Helstrom bound for squeezed states (SQZHEL).}
    \label{fig:2}
\end{figure*}

\mbox{Figure 2 (a)} shows the measured probability of error using displaced squeezed states. In each run of the experiment the squeezing level is fixed and the mean photon number of the state varies solely with the amplitude of the displacement. As there is a minimum amount of energy in the state due to the squeezing, the traces modelling the experiment do not begin at $\bar{n}=0$. The limits for the probability of error using different protocols are shown for comparison, and we clearly demonstrate discrimination of the S-BPSK signal using homodyne detection with lower errors than possible for a protocol using a C-BPSK signal and an optimal quantum receiver. Notably, we demonstrate an error probability $3.28$ standard deviations below the coherent Helstrom bound using only Gaussian resources.

In principle the experimental traces should be tangent to the Gaussian limit when they have an optimal squeezing ratio, however due to impurities in the squeezing, and the presence of electronic noise, they are not. (See the appendices for more details.) In \mbox{Figure 2 (b)} the error probabilities for three S-BPSK signals with approximately the same energy but with different levels of squeezing are shown. Even for very weak signals there is always some advantage to be gained through squeezing, consequently there is always a S-BPSK signal that can outperform the SQL for any $\bar{n}$. However in order to beat the coherent Helstrom bound the signal energy must be above a certain value, found by numerical evaluation of the coherent Helstrom and Gaussian bounds to be $\bar{n} \approx 0.67$.

\begin{figure}
    \centering
    \includegraphics[width=\linewidth]{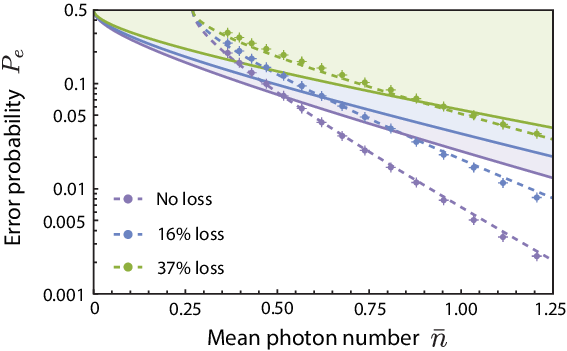}
    \caption{Error probability as a function of mean photon number for different added channel loss. Markers show measured values using a S-BPSK signal with 3.8 dB squeezing. Dashed curves are our model of the experiment, and solid curves show the C-BPSK limits. The traces labelled \textquote{no loss} refer to no added channel loss; the intrinsic loss of the detection system is still present. Error bars are obscured by the marker size.}
    \label{fig:3}
\end{figure}

Figure 3 shows the performance of the same S-BPSK signals for three channels with different added loss. The ideal performance of equivalent energy C-BPSK signals are shown as a comparison. It is clear that the performance of the S-BPSK signal degrades faster than in the C-BPSK case, as there are some states that no longer surpass the coherent state limit when loss is present. There is in principle always some improvement of the SNR that can be obtained regardless of the channel loss, however the improvement may not be worth the cost of allocating signal energy to squeezing.

The performance of squeezed states degrades rapidly when optical losses are present in the channel. The adjusted error rate model for a channel with loss $L\in[0,1]$ is
\begin{align}
    P_e = \frac{1}{2}\text{Erfc}[\sqrt{\frac{2(\bar{n}V^2-\sinh^2(r)-(v_{th}-1)/2)(1-L)}{(e^{-2r}v_{th}(1-L)+L+v_{en})}}].
\end{align}
\begin{figure}
    \centering
    \includegraphics[width=\linewidth]{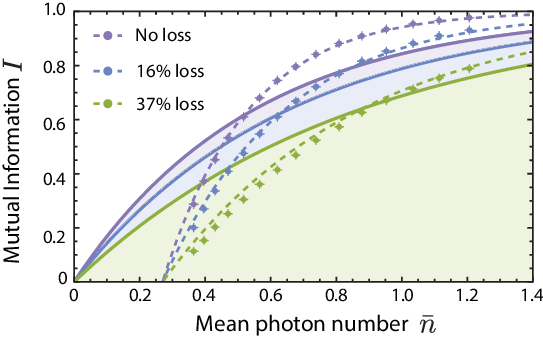}
    \caption{Mutual information as a function of mean photon number for different added channel loss. Markers show values calculated from the measured probability of error using a S-BPSK signal with 3.8 dB squeezing. Dashed curves are our model of the experiment, and solid curves show the C-BPSK limits. The traces labelled \textquote{no loss} refer to no added channel loss; the intrinsic loss of the detection system is still present. Error bars are obscured by the marker size.}
    \label{fig:4}
\end{figure}

An alternative figure of merit to the probability of error is the mutual information
\begin{align}
    I = 1-P_e\log_2(\frac{1}{P_e})-(1-P_e)\log_2(\frac{1}{1-P_e})
\end{align}
between the transmitter and receiver. In out case due to the symmetry of the protocol this is equivalent to the capacity of a binary-symmetric channel, and has a theoretical maximum of one bit per use when $P_e\rightarrow0$. \mbox{Figure 4} shows the mutual information calculated from the error rates of the S-BPSK signal in lossy channels. We can consider the introduction of squeezing to be a modification of the channel itself \cite{PhysRevA.64.014304}; specifically we are lowering the additive noise of the channel. The channel capacity is an asymptotic limit that can be approached by using error correction codes, and due to the lower probability of error the S-BPSK signal could potentially achieve the same data rates as C-BPSK signals while using shorter error correction codes.

\noindent {\it Discussion.—}The introduction of quadrature squeezing greatly enhances the SNR of BPSK signals, allowing the coherent Helstrom bound to be passed with only readily-available Gaussian optics. We have demonstrated this experimentally for the first time by generating quantum states with an optimised ratio of squeezing to displacement. The impact of such states for quantum enhanced communication has also been analysed.

On a practical level there remain several issues that limit the immediate applicability of our results. Realistic communications systems are almost always limited by channel losses, and the improvement from squeezing can quickly be diminished, as shown in \mbox{Figure~\ref{fig:3}}. Additionally, larger signal alphabets are required to attain higher channel capacity, so it would be desirable to extend the BPSK signal alphabet to higher-order amplitude-shift-keyed (ASK) or phase-shift-keyed (PSK) alphabets. Because ASK signals can always be decoded by measuring only a single quadrature we predict a similar improvement from using squeezed states instead of coherent states, as is found for BPSK. However PSK signals, which classically are more energy efficient, require both quadratures to be measured when there are more than two signal states. Doing so involves splitting the received signal in half and consequently incurring a 3 dB loss to the signal amplitude. For squeezed states the SNR penalty is even greater as vacuum noise is coupled into the squeezed quadrature. 
Consequently we find that for M-PSK with $M>2$ there is little to no advantage to be obtained from using squeezed signals and dual homodyne detection \footnote{A more detailed discussion can be found in the appendices.}. 
For the specific case of 4-PSK coherent states are optimal, and squeezing can only degrade the performance of the signal. In fact, better performance can be obtained with a 4-ASK signal, in contrast with the classical case. For larger alphabets the optimal Gaussian signal is a displaced squeezed state with a small amount of phase-squeezing \cite{PhysRevA.86.022306}. \par
Despite these issues, we believe that the higher distinguishability of displaced squeezed states is of fundamental interest. The advantage offered by squeezed states may translate to other types of quantum information processing where the energy of signal states is constrained and optical losses can be kept to a minimum.

Finally, we wish to draw a comparison with quantum sensing. For many years, squeezed states of light have been used to enhance the sensitivity of quantum measuring devices, for example, in gravitational wave detection \cite{LIGO,oelker2014squeezed}, in achieving Heisenberg-limited sensing \cite{PhysRevLett.100.073601}, and in other related tasks \cite{roga2015device,bradshaw2021optimal}. We believe we have demonstrated an analogous advantage for state discrimination in this work.

\noindent {\it Acknowledgements—}This research was funded by the Australian Research Council Centre of Excellence CE170100012.
This research was also supported by A*STAR C230917010, Emerging Technology and A*STAR C230917004, Quantum Sensing.

\noindent {\it Data availability—}The data is available from the authors upon reasonable request.

\bibliography{statedisc}


\clearpage
\onecolumngrid
\begin{center}
  \textbf{\large Appendices for\\[3pt]
  \textquote{All-Gaussian State Discrimination Beyond the Coherent Helstrom Bound}}
\end{center}
\vspace{10pt}
\setcounter{section}{0}
\setcounter{figure}{0}
\setcounter{table}{0}
\renewcommand{\thesection}{A\arabic{section}}
\renewcommand{\thefigure}{A\arabic{figure}}
\renewcommand{\thetable}{A\arabic{table}}

\section{Derivation of the optimal Gaussian protocol}\label{sec:derivation}
Demonstrating that displaced squeezed states followed by homodyne detection forms the optimal Gaussian scheme for discriminating a binary-phase-shift-keyed (BPSK) signal requires two steps. First, subject to the constraint of fixed mean photon number $\bar{n}$, we maximise the signal-to-noise ratio (SNR) of an arbitrary Gaussian state and show that this results in a displaced squeezed state. We then demonstrate that after the signal is defined, but prior to its measurement by a homodyne receiver, no Gaussian operations can improve the SNR. It's worth emphasising a subtlety: applying a squeezing operation to a coherent signal does not change the SNR, however it does change the mean energy of the state, and compared to a coherent state with the changed energy, the displaced squeezed state may have better SNR. \par
An arbitrary Gaussian state can be constructed as follows: starting with a thermal state $\hat{\rho}_{th}$ with mean photon number $n_{th}$ we apply a squeezing operation $\hat{S}(r)$. We assume $r$ is real apply a rotation $\hat{R}(\theta)$ to obtain squeezing in an arbitrary quadrature. Finally we apply a displacement $\hat{D}(\alpha)$ to obtain the state
\begin{align}
    \hat{\rho}_0 = \hat{D}(\alpha)\hat{R}(\theta)\hat{S}(r)\hat{\rho}_{th}\hat{S}^\dagger(r)\hat{R}^\dagger(\theta)\hat{D}^\dagger(\alpha).
    \label{rhogen}
\end{align}
It can be assumed without loss of generality that $\alpha$ is real, as changing the angle between squeezing and displacement can be done with $\theta$, and a rotation of the entire state can be achieved by measuring a rotated quadrature operator. Using the convention $\hat{X} = (\hat{a}+\hat{a}^\dagger)/2$, the first two moments of the measured quadrature are
\begin{align}
    \braket{\hat{X}} &= \alpha, \\
    \braket{\Delta\hat{X}^2} &= \frac{2n_{th}+1}{4}\left(e^{-2r}\cos^2(\theta)+e^{2r}\sin^2(\theta)\right),
\end{align}
so that the homodyne SNR 
\begin{equation}
    \text{SNR} =\frac{\braket{\hat{X}}^2}{\braket{\Delta\hat{X}^2}}    
\end{equation}
of the state is 
\begin{align}
    \text{SNR} = \frac{4\alpha^2}{(2n_{th}+1)(e^{-2r}\cos^2(\theta)+e^{2r}\sin^2(\theta))}.
    \label{snrgen}
\end{align}
The mean photon number of the state in Equation (\ref{rhogen}) is
\begin{align}
    \bar{n} = \alpha^2 + n_{th} + (2n_{th}+1)\sinh^2(r),
    \label{ngen}
\end{align}
so for a state of fixed energy any thermal occupation both inflates the denominator of Equation (\ref{snrgen}) and consumes energy that would otherwise be available for displacement. It is therefore clear by inspection that the SNR is maximised when $\theta = 0$ and $n_{th}=0$: the optimal Gaussian signal state is a pure displaced squeezed state with its squeezed quadrature aligned to the displacement.
\par Setting $\theta = n_{th} = 0$, the energy constraint of Equation (\ref{ngen}) reduces to $\bar{n} = \alpha^2+\bar{n}_s$, where $\bar{n}_s = \sinh^2(r)$ is the mean photon number contribution from squeezing, and the SNR becomes $\text{SNR} = 4(\bar{n}-\bar{n}_s)e^{2r}$. Writing $e^{r} = \cosh(r)+\sinh(r) = \sqrt{\bar{n}_s+1}+\sqrt{\bar{n}_s}$ this is
\begin{align}
    \text{SNR} = 4(\bar{n}-\bar{n}_s)(\sqrt{\bar{n}_s}+\sqrt{\bar{n}_s+1})^2.
    \label{snropt}
\end{align}
It is most convenient to carry out the optimisation over $r$ directly. Differentiating $\text{SNR}(r) = 4(\bar{n}-\sinh^2(r))e^{2r}$ gives
\begin{align}
    \frac{d\,\text{SNR}}{dr} = 8e^{2r}\left(\bar{n}-\sinh^2(r)-\sinh(r)\cosh(r)\right),
\end{align}
so the stationary point satisfies the compact condition
\begin{align}
    \bar{n} = \sinh(r)\left(\sinh(r)+\cosh(r)\right) = e^{r}\sinh(r).
    \label{optcond}
\end{align}
Equation (\ref{optcond}) is solved by $\bar{n}_s = \sinh^2(r) = \bar{n}^2/(2\bar{n}+1)$, equivalently $e^{2r} = 2\bar{n}+1$, so the optimal squeezing ratio is $\gamma = \bar{n}_s/\bar{n} = \bar{n}/(2\bar{n}+1)$. Substituting back, $\bar{n}-\bar{n}_s = (\bar{n}^2+\bar{n})/(2\bar{n}+1)$, and the maximum SNR attainable by any Gaussian state of mean photon number $\bar{n}$ is
\begin{align}
    \text{SNR}_{\text{max}} = 4(\bar{n}^2+\bar{n})
    \label{snrmax}
\end{align}
\cite{PhysRevA.97.032315}. Two limits are worth noting. For $\bar{n}\ll1$ we have $\gamma\rightarrow\bar{n}$ and $\text{SNR}_{\text{max}}\rightarrow4\bar{n}$, recovering the coherent state result, so the Gaussian advantage vanishes for very weak signals. For $\bar{n}\gg1$ we have $\gamma\rightarrow1/2$, and the energy is shared equally between squeezing and displacement.
\par The same calculation can be carried through for the impure squeezing that is actually available in the laboratory. Writing $v_{th} = 2n_{th}+1$ for the thermal noise of the source relative to shot noise, as in the main text, Equations (\ref{snrgen}) and (\ref{ngen}) give $\text{SNR} = 4\alpha^2e^{2r}/v_{th}$ subject to $\bar{n} = \alpha^2+(v_{th}-1)/2+v_{th}\sinh^2(r)$, and the stationarity condition (\ref{optcond}) generalises to
\begin{align}
    e^{r}\sinh(r) = \frac{\bar{n}-(v_{th}-1)/2}{v_{th}}.
\end{align}
Excess thermal noise therefore reduces the optimal squeezing ratio and lowers the attainable SNR below Equation (\ref{snrmax}). This, together with the below-unity homodyne visibility, is the origin of the gap between our data and the Gaussian limit, which is examined in \mbox{Section~\ref{sec:imperfections}}.
\par We can now show that any Gaussian unitary applied prior to homodyne detection does not decrease the probability of error. It is trivial to see that rotating the received states by an angle $\phi$ can only be detrimental, as not only will the amplitude decrease but noise from the anti-squeezed quadrature will be introduced:
\begin{align}
    \text{SNR}(\phi) = \frac{4\alpha^2\cos^2(\phi)}{e^{-2r}\cos^2(\phi)+e^{2r}\sin^2(\phi)}.
\end{align}
Applying further squeezing will change both the signal and the noise by a factor of $e^{-2r'}$ resulting in no net change to the SNR. Finally we note that while displacement by an amplitude $\alpha'$ will increase the SNR (for one of the signal states), in order for the protocol to remain symmetric the decision threshold will also have to shift from $0\rightarrow\alpha'$, consequently the error probability is unchanged. As any Gaussian unitary can be decomposed into the above transformations, these considerations encompass the totality of Gaussian operations that may be performed on the state.
\par Therefore we can conclude that a displaced squeezed state signal with $\gamma =\bar{n}/(2\bar{n}+1)$ followed by homodyne detection constitutes the optimal scheme for all-Gaussian BPSK in loss-free channels. (In a lossy channel the above argument still holds, however the optimal squeezing ratio $\gamma$ will be reduced.) Finally we note that because the SNR is unchanged by any squeezing operation after transmission, any receiver that reaches the coherent Helstrom bound can be made to reach the squeezed state Helstrom bound 
\begin{align}
    P = \frac{1}{2}(1-\sqrt{1-e^{-4(\bar{n}^2+\bar{n})}}),
\end{align}
simply by the inclusion of an anti-squeezing operation that transforms the incoming squeezed states into coherent states.

\section{Experimental Imperfections}\label{sec:imperfections}

\begin{figure}[h]
    \centering
    \includegraphics[width=0.75\linewidth]{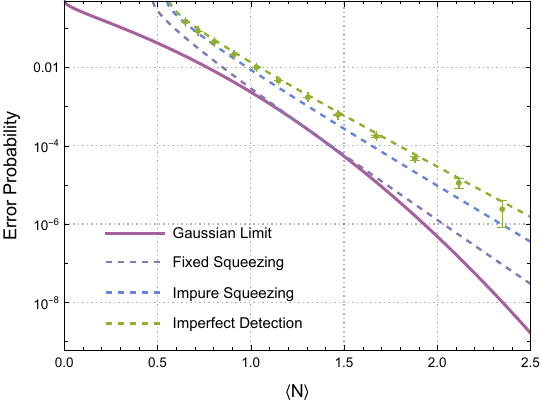}
    \caption{Experimental probability of error for a 5 dB squeezed BPSK signal and homodyne detection, against the mean photon number of the signal. The solid line shows the Gaussian limit, and the error probability is modelled with dashed lines, each accounting for one additional imperfection.}
    \label{fig:noisebreak}
\end{figure}

As can be seen in \mbox{Figure~\ref{fig:2}}, we are unable to saturate the Gaussian limit for the displaced squeezed state signal. This is due to the practical constraints on our system, as outlined in the derivation of \mbox{Equation~\ref{expmod}}. How each of the parameters in that expression contribute to the error rate is shown in \mbox{Figure~\ref{fig:noisebreak}}. First, as mentioned in the main text, by setting the squeezing to a fixed level we introduce a minimum state energy. The resulting error rate, depicted as the purple dashed curve, only saturates the Gaussian limit for a particular value of the total energy. However, the squeezed light generated by an OPO is not pure and carries excess thermal energy. This increases the minimum energy of the state, and also degrades the scaling of the error probability, as seen in the dashed blue curve. Finally, the below-unity visibility further degrades the performance of the protocol, and its inclusion in our model gives us an accurate fit to the data. Note that as detector inefficiency is equivalent to loss, this must be accounted for in calculating the total energy of the state, hence its appearance in \mbox{Equation~\ref{etot}}. In contrast, the reported values of squeezing are the measured values, and are not inferred. We did not find it necessary to include phase noise in our analysis owing to both the stability of our system and the robustness of homodyne detection to this kind of noise.

\section{Three state discrimination}
\subsection{Amplitude shift keying}
The three state amplitude shift keyed signal is
\begin{equation}
    \ket{\psi_k} = \hat{D}(k\alpha)\hat{S}(r)\ket{0}, k\in\{-1,0,1\},
\end{equation}
which results in three squeezed states in a line along the amplitude quadrature. As only two signal states have coherent displacements the energy allocation is uneven. The average signal energy is 
\begin{equation}
    \bar{n} = \frac{2}{3}\alpha^2+\sinh^2(r).
\end{equation}
Again we set $\gamma = \sinh^2(r)/\bar{n}$ leading to the amplitude of the displacement being
\begin{align}
    \alpha = \sqrt{\frac{3}{2}(1-\gamma)\bar{n}}
\end{align}
All three symbols are equally likely and equally spaced, so under an optimal decision strategy the homodyne outcome $x$ is assigned to the nearest signal state and the decision thresholds lie at $x = \pm\alpha/2$. Each state has quadrature variance $\braket{\Delta\hat{X}^2} = e^{-2r}/4$, so the error rates follow from integrating the homodyne distributions outside the relevant decision region. The inner $k=0$ state is misidentified whenever $|x|>\alpha/2$, whereas each of the outer $k=\pm1$ states can only be misidentified on one side:
\begin{align}
    P(\text{err}|k=0) &= \text{Erfc}\left[\frac{e^r\alpha}{\sqrt{2}}\right], \\
    P(\text{err}|k=\pm1) &= \frac{1}{2}\text{Erfc}\left[\frac{e^r\alpha}{\sqrt{2}}\right].
\end{align}
Averaging over the three symbols, the average probability of error is
\begin{equation}
    P = \frac{2}{3}\text{Erfc}[\frac{e^r\alpha}{\sqrt{2}}].
\end{equation}
For coherent states, where $\gamma=0$ and $\alpha = \sqrt{3\bar{n}/2}$, this reduces to 
\begin{equation}
    P_c = \frac{2}{3}\text{Erfc}[\frac{\sqrt{3\bar{n}}}{2}].
\end{equation}
For squeezed states the quantity to be maximised is $\alpha^2e^{2r} = \frac{3}{2}(\bar{n}-\bar{n}_s)e^{2r}$, which differs from the binary case of Equation (\ref{snropt}) only by the constant factor $3/2$. The optimisation is therefore identical to that of \mbox{Section~\ref{sec:derivation}}, and we once again find the optimal splitting ratio to be $\gamma = \bar{n}/(2\bar{n}+1)$. This gives a probability of error
\begin{equation}
    P_s= \frac{2}{3}\text{Erfc}[\frac{\sqrt{3(\bar{n}^2+\bar{n})}}{2}],
\end{equation}
which shows the same quadratic scaling in the argument of the error function as is found for binary signals, albeit with a smaller prefactor arising from the uneven energy allocation, since only two of the three symbols carry a displacement. Note that the uneven allocation changes the attainable error rate but not the optimal distribution of energy between squeezing and displacement.
\subsection{Phase shift keying}
The three state phase shift keyed signal is
\begin{equation}
    \ket{\psi_k} = \hat{R}(k\frac{2\pi}{3})\hat{D}(\beta)\hat{S}(r)\ket{0}, k\in\{-1,0,1\}.
\end{equation}
This divides phase space into three wedges and the error rate is the probability that the result of a dual-homodyne measurement is outside the appropriate wedge. By symmetry we can consider just the error probability of the $k=0$ state that is both displaced and squeezed along the amplitude quadrature. The joint probability density function for the dual-homodyne measurement is then
\begin{align}
    P(x,p) = \mathcal{N}(x;\frac{\alpha}{\sqrt{2}},\frac{1+e^{-2r}}{8})\times\mathcal{N}(p;0,\frac{1+e^{2r}}{8}),
\end{align}
where $\mathcal{N}(z;\mu_z,\sigma_z^2)$ is a normal distribution. Note that we are using the convention that $\hat{X}=(\hat{a}+\hat{a}^\dagger)/2$. We can see that both the amplitude has been reduced and vacuum noise has been introduced, due to the necessity of a balanced beam-splitter in the dual-homodyne receiver. The one-third of phase space that we need to integrate over is defined by the bounds $p=\tan(\pm\pi/3)x$, which translates to the constraints $\sqrt{3}x>|p|$ and $x>0$, so the probability of error is given by the integral
\begin{align}
    P = 1-\int_0^\infty dx\int_{-\sqrt{3}x}^{\sqrt{3}x}dp P(x,p),
\end{align}
for which we do not have an analytic expression. Alternatively, the probability distribution over $x$ and $p$ can be cast in terms of the phase angle $\theta = \arctan(\frac{p}{x})$; Aalo et al. \cite{4400760} have shown that the resulting distribution is
\begin{equation}
    P(\theta) = \frac{e^{-\mu_x^2/2\sigma_x^2}}{4\pi\sigma_x\sigma_p\mathcal{A}(\theta)}(\frac{\sqrt{\pi}\mathcal{B}(\theta)}{2\sqrt{\mathcal{A}(\theta)}}*e^{\frac{\mathcal{B}(\theta)^2}{4\mathcal{A}(\theta)}}*\text{Erfc}[\frac{-\mathcal{B}(\theta)}{2\sqrt{\mathcal{A}(\theta)}}]+1),
\end{equation}
where 
\begin{equation}
    \mathcal{A}(\theta) = \frac{\cos^2(\theta)}{2\sigma^2_x}+\frac{\sin^2(\theta)}{2\sigma^2_p}
\end{equation}
and 
\begin{equation}
    \mathcal{B}(\theta) = \frac{\mu_x\cos(\theta)}{\sigma^2_x}.
\end{equation}
\par
\begin{figure}
    \centering
    \includegraphics[width=0.75\linewidth]{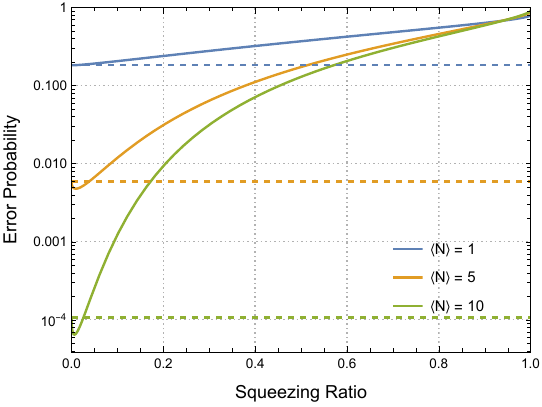}
    \caption{Probability of error for the Ternary Phase-Shift Keying signal, as a function of the proportion of energy allocated to squeezing, for different values of the total state energy. Solid lines are calculated by numerically integrating over the phase space coordinates, and dashed lines show the coherent state error.}
    \label{fig:enter-labe}
\end{figure}
\begin{figure}
    \centering
    \includegraphics[width=0.75\linewidth]{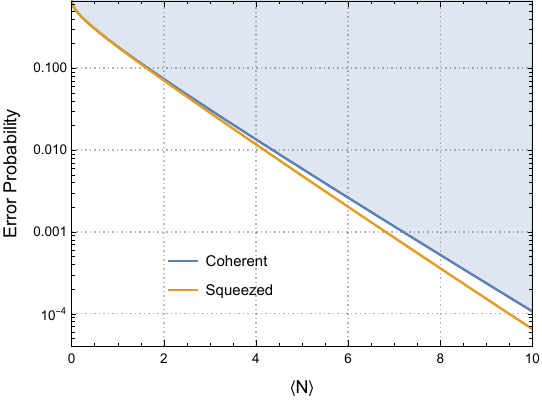}
    \caption{Probability of error for the Ternary Phase-Shift Keying signal, as a function of the total state energy, for coherent states (blue) and displaced squeezed states with $0.5\%$ of state energy in squeezing.}
    \label{fig:enter-label}
\end{figure}

Regardless of which representation we use, we must turn to numerical integration to find the probability of error. By integrating $P(\theta)$ over the range $\theta\in[-\pi/3,\pi/3]$ for different values of the squeezing ratio and state energy we find that only a slight improvement in the error rate is possible. As seen in \mbox{Figure~\ref{fig:enter-labe}}, the largest improvement appears to be at a constant squeezing ratio of approximately $0.5\%$. Fixing the ratio to this value and varying the total energy of the signal demonstrates how marginal the improvement offered by squeezing is, as shown in Figure~\ref{fig:enter-label}.
 
\section{Four state discrimination}
The quadrature phase shift keyed (QPSK) signal is 
\begin{equation}
    \ket{\psi_k}=\hat{R}(\frac{\pi(2k+1)}{4})\hat{D}(\beta)\hat{S}(r)\ket{0},k\in\{0,1,2,3\}.
\end{equation}
The additional rotation by $\pi/4$ places the decision boundaries of the dual-homodyne receiver along the $x$ and $p$ axes, so each symbol can be decoded from the sign of the two quadratures independently. Projected onto those axes the states have symmetric variance in both quadratures
\begin{equation}
    V = (e^{2r}+e^{-2r})/8 = \cosh(2r)/4,
\end{equation}
which is minimised for $r=0$. As the coherent amplitude of a state with fixed energy is maximised for $r=0$ squeezing does not improve the performance of this signal.
\par
\begin{figure}
    \centering
    \includegraphics[width=0.75\linewidth]{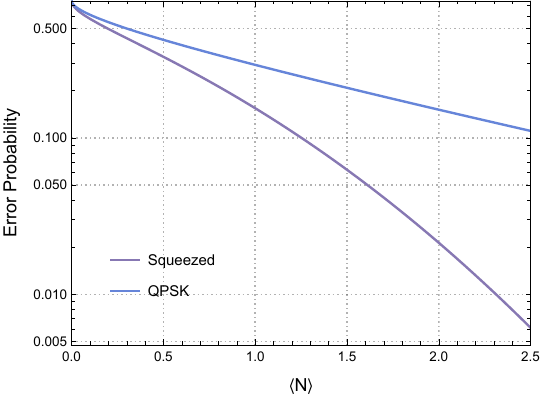}
    \caption{Probability of error of the optimal QPSK and 4-ASK signals, as a function of the total state energy. For QPSK coherent states are optimal, whereas for ASK squeezing provides improved performance.}
    \label{fig:4statesP}
\end{figure}

\begin{figure}
    \centering
    \includegraphics[width=0.75\linewidth]{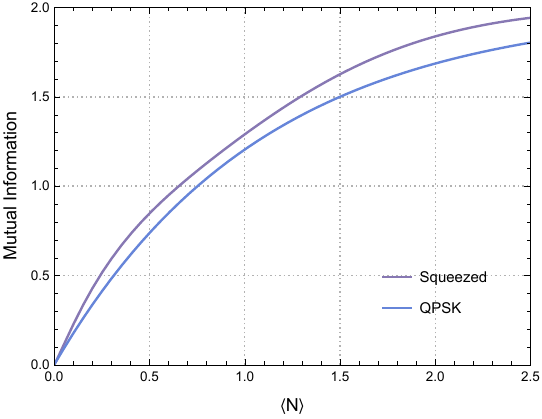}
    \caption{Mutual information of the optimal QPSK and 4-ASK signals, as a function of the total state energy. Quadrature squeezing allows the ASK signal to more quickly saturate the channel capacity of 2 bits per use.}
    \label{fig:4stateMI}
\end{figure}

 This can be made quantitative. The balanced beam splitter of the dual-homodyne receiver quarters the measured signal power and couples in an equal amount of vacuum noise, so for the $k=0$ state the measured moments are $\braket{\hat{X}} = \beta/2$ and $\braket{\Delta\hat{X}^2} = (\cosh(2r)+1)/8$. Using $\beta^2 = \bar{n}(1-\gamma)$ and $\sinh^2(r) = \gamma\bar{n}$, so that $\cosh(2r) = 1+2\gamma\bar{n}$, the per-quadrature SNR of the signal is 
\begin{equation}
    \text{SNR}=\frac{\bar{n}(1-\gamma)}{\bar{n}\gamma+1},
\end{equation}
and
\begin{equation}
    \frac{d\,\text{SNR}}{d\gamma} = -\frac{\bar{n}(1+\bar{n})}{(1+\gamma\bar{n})^2}<0
\end{equation}
for all $\bar{n}>0$. The SNR is therefore maximised for $\gamma=0$ and falls monotonically as energy is moved into squeezing. For quaternary PSK read out by dual homodyne detection the optimal Gaussian signal is thus a coherent state, in contrast to the binary and amplitude shift keyed cases, and the same conclusion applies to the higher-order PSK alphabets for which both quadratures must likewise be measured.
The optimal QPSK probability of error is therefore
\begin{align}
    P &= 1-\frac{1}{4}\bigg(1+\text{Erf}\Big[\sqrt{\frac{\bar{n}}{2}}\Big]\bigg)^2.
\end{align}
In this expression the argument of the error function scales as $\sqrt{\bar{n}}$, however from the binary and ternary cases we expect that for a 4-ASK signal this scaling would be $\bar{n}$. Following the same steps as before we find that the optimal probability of error for the 4-ASK signal is 
\begin{align}
    P_s = \frac{3}{4}\text{Erfc}\bigg[\sqrt{\frac{2}{5}}\sqrt{\bar{n}^2+\bar{n}}\bigg].
\end{align}
We compare these two error rates in Figure~\ref{fig:4statesP}, where it is seen that the squeezed state ASK signal achieves a lower error. However, it should be noted that this is a symbol error and not a bit error, and that the bit error for each signal will depend on how digital information is encoded in each symbol. To achieve a comparison that avoids these complications we calculate the mutual information of each signal, as seen in Figure~\ref{fig:4stateMI}.  

\end{document}